\input{aipcheck}
\documentclass[
    ,final            
  ]
  {aipproc}

\layoutstyle{6x9}
\newcommand{\teff}{T_{\rm eff}}

\begin{document}

\title{$\teff$ and $\log{g}$ dependence of FeH in M-dwarfs}


\author{S.~Wende}{
  address={Institut f\"ur Astrophysik, Georg-August-Universit\"at
  G\"ottingen, Friedrich-Hund Platz 1, D-37077 G\"ottingen, Germany}
}

\author{A.~Reiners}{
  address={Institut f\"ur Astrophysik, Georg-August-Universit\"at
  G\"ottingen, Friedrich-Hund Platz 1, D-37077 G\"ottingen, Germany}
}

\author{H.-G.~Ludwig}{
  address={GEPI, CIFIST, Observatoire de Paris-Meudon, 5 place Jules Janssen,
  92195 Meudon Cedex, France}
}

\begin{abstract}
 We present synthetic FeH band spectra in the $z$-filter range for several
 M-dwarf models with $\log{g}=3.0-5.0$~[cgs] and $\teff=2800~{\rm K}-3450~{\rm
 K}$. Our aim is to characterize convective velocities in M-dwarfs and to give
 a rough estimate of the range in which 3D-atmosphere treatment is necessary
 and where 1D-atmosphere models suffice for the interpretation of
   molecular spectral features. This is also important in order to
 distinguish between the velocity-broadening and the rotational- or
 Zeeman-broadening. The synthetic spectra were calculated using 3D
 \verb!CO5BOLD! radiative-hydrodynamic (RHD) models and the line synthesis
 code \verb!LINFOR3D!. We used complete 3D-models and high resolution 3D
 spectral synthesis for the detailed study of some well isolated FeH
 lines. The FeH line strength shows a dependence on surface gravity and
 effective temperature and could be employed to measure both quantities in
 M-type objects. The line width is related to the velocity-field in the model
 stars, which depends strongly on surface gravity. Furthermore, we investigate
 the velocity-field in the 3D M-dwarf models together with the related micro-
 and macro-turbulent velocities in the 1D case. We also search for effects on
 the lineshapes.

\end{abstract}

\maketitle


\section{Single 3D- and <3D>- ${\bf FeH}$ Lines}

\begin{figure}[!h]
\resizebox{0.9\textwidth}{!}{
  \includegraphics{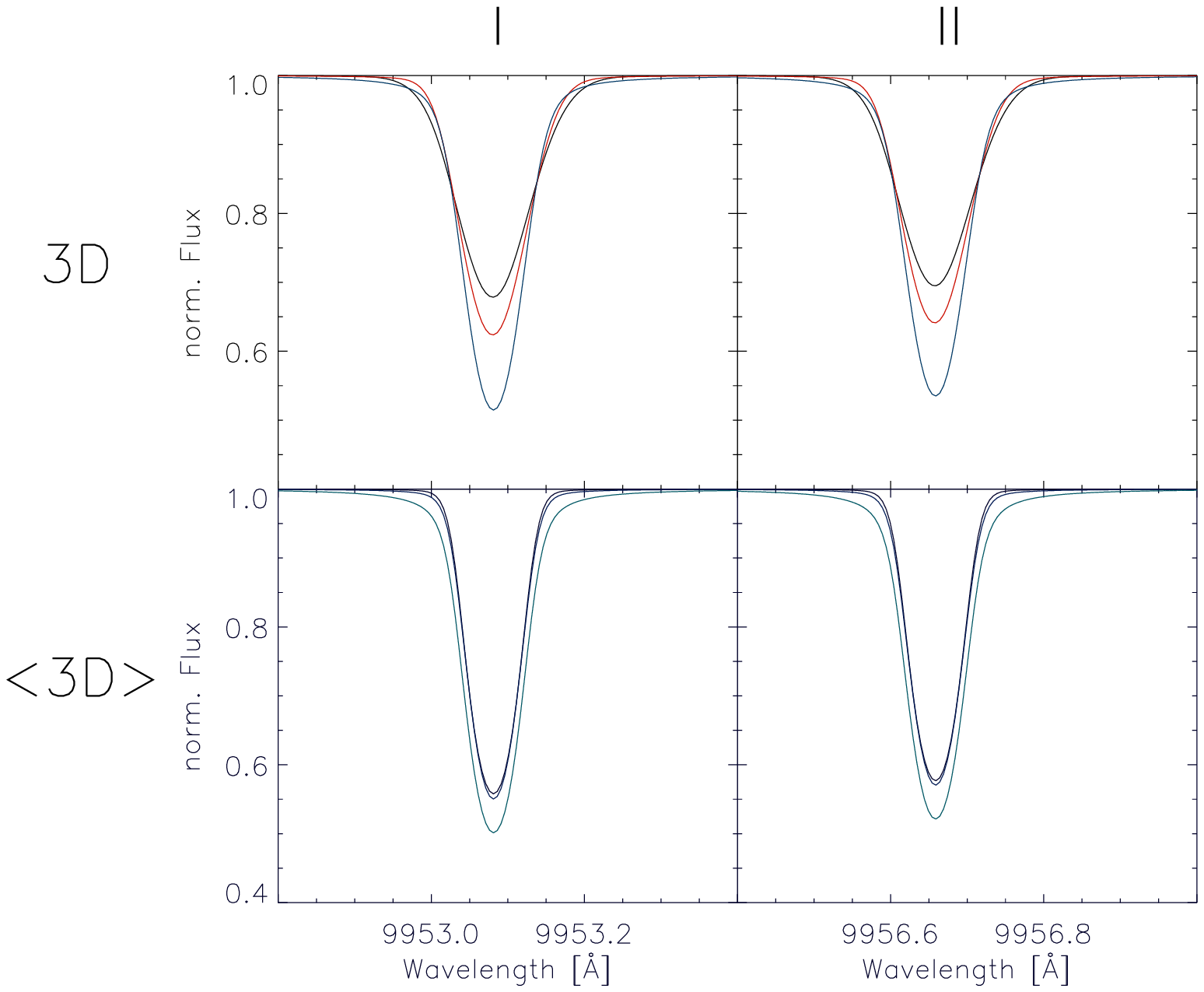}
  \includegraphics{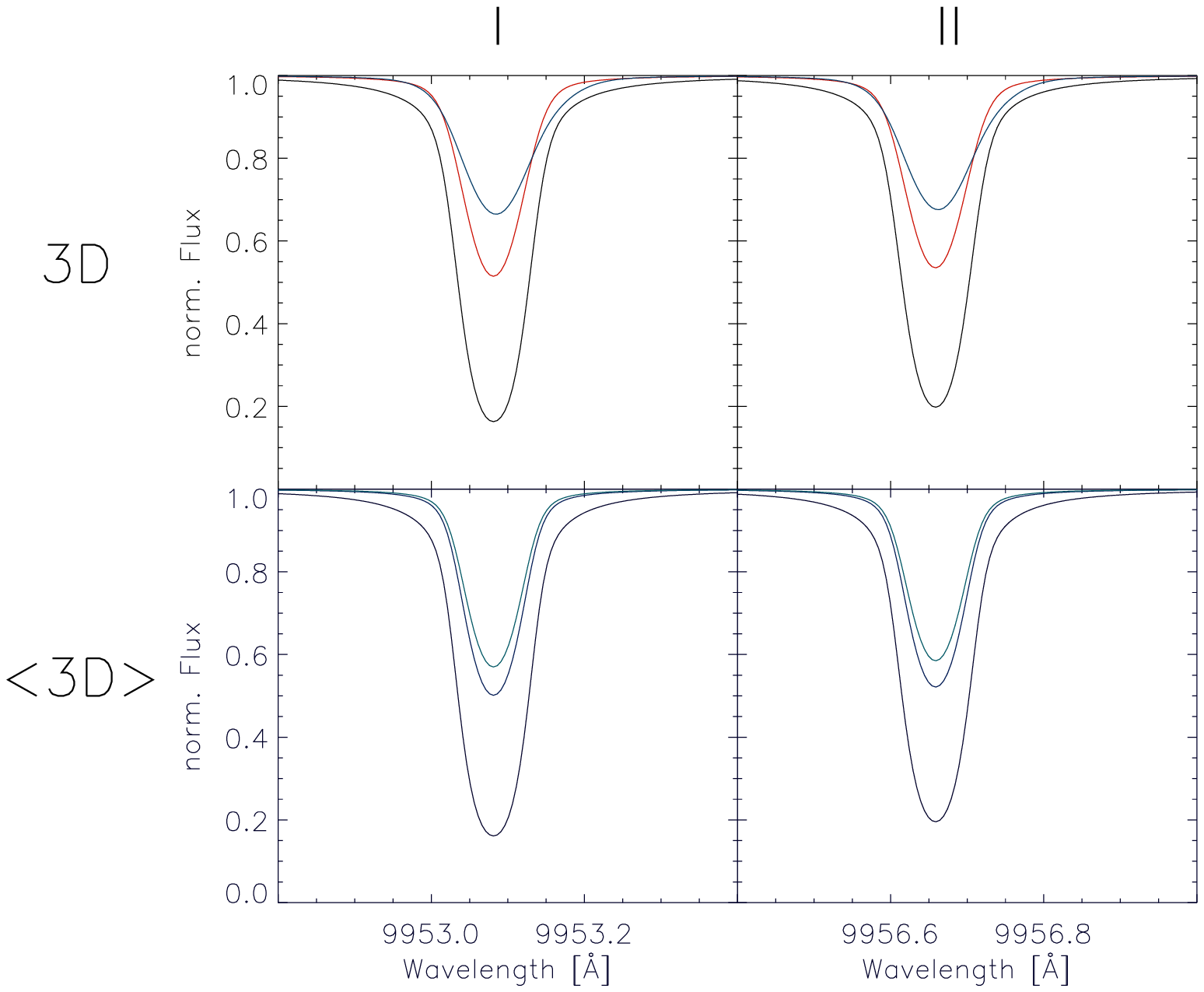}}
 \caption{\emph{Left}: Two out of ten simulated FeH lines (Tab.~\ref{tab2})
  for models with $\teff=3300K$ and $\log{g}$ values of 3.0, 4.0, and 5.0
  [cgs]. The line centered at vacuum wavelength $9953.07$~\AA\ is magnetically
  sensitive (label I) and the line centered at vacuum wavelength
  $9956.72$~\AA\ is magnetically insensitive (label II). The upper panels show
  3D lines with $\log{g}$ = 5.0 (blue), 4.0 (red), and 3.0 (black) [cgs]. The
  lower panels show the same for the <3D> lines. \emph{Right}: The same as in
  the left plot for models with $\teff$ of $2800~{\rm K}$, $3300~{\rm K}$ and
  $3450~{\rm K}$. All models have $\log{g}=5.0$ [cgs].  The upper panels show
  3D lines with $\teff$ = $3450~{\rm K}$ (blue), $3300~{\rm K}$ (red), and
  $2800~{\rm K}$ (black). The lower panels show the same for the <3D> lines.}
\label{fig1}
\end{figure}

In this work, we use the radiative-hydrodynamic (RHD) model-atmospheres (see
Tab.~\ref{tab1}) which were computed with the \verb!CO5BOLD!-code employing
opacity binning (see e.g.~\cite{2006A&A...459..599L} or
\cite{2002A&A...395...99L}) with five bands to take into account the
wavelength dependence of the radiative energy exchange.  To calculate the
spectral lines, we use the line synthesis code
\verb!LINFOR3D!\footnote{see
    \url{http://www.aip.de/~mst/Linfor3D/linfor\_3D\_manual.pdf}}, which is
 able to use the 3D capacity of the
\verb!CO5BOLD!-atmosphere models. We investigated 10 FeH lines between
$9950$~\AA\ and $9990$~\AA\ which were chosen from \cite{2006ApJ...644..497R}
(see Tab.~\ref{tab2}). The resolution of the synthetic spectra is $\Delta\lambda =
5\cdot 10^{-3}$\AA\ corresponding to a Doppler velocity of $\approx$150~m/s at
the wavelength of the considered lines. We averaged over the spatial x-y
directions in the 3D-model to receive a 1D-model, which we call the
<3D>-model. This model is used for comparison between the modelled 3D and <3D> FeH
lines to investigate the effect of micro and macro turbulent velocities. The
3D velocity fields are not taken into account in the <3D>-models, and line
broadening is treated within the classical scheme of micro- and
macro-turbulence. In Fig.~\ref{fig1} one can see that there is a strong
influence of surface gravity and effective temperature on the lineshape for
the 3D-models. Regarding the <3D>-models, the effect on line depth with
changing $\log{g}$ is less obvious. Line depth changes in the 3D case can be
explained by line broadening by the hydrodynamical velocity field. The
equivalent width of the lines is not strongly affected since they are only
mildly saturated (see Fig.~2).

 \begin{table}[!h]
\begin{tabular}{lrrrrr}
\hline
  \tablehead{1}{l}{b}{Model code}
& \tablehead{1}{r}{b}{Dim.}
& \tablehead{1}{r}{b}{Size [km]} 
& \tablehead{1}{r}{b}{Opacities}
& \tablehead{1}{r}{b}{$\teff$ [K]}
& \tablehead{1}{r}{b}{$\log{g}$} \\
\hline
d3t33g30mm00w1 & 3 & 85000 x 85000 x 58350 & PHOENIX & 3300 & 3.0 \\ 
3dt3280g40mm00 & 3 & 4500 x 4500 x 1850 & PHOENIX & 3300 & 4.0 \\ 
d3t33g50mm00w1 & 3 & 300 x 300 x 260 & PHOENIX & 3300 & 5.0 \\
d3t35g50mm00w1 & 3 & 900 x 900 x 300  & PHOENIX & 3450 & 5.0 \\ 
d3t28g50mm00w1 & 3 & 250 x 250 x 270 & PHOENIX & 2800 & 5.0 \\ 
\hline
\end{tabular}
\caption{Overview of different model quantities}
\label{tab1}
\end{table}

%
%
\begin{table}[!h]
\begin{tabular}{crrrrcrrrr}
\hline
\tablehead{1}{l}{b}{$\lambda_{\rm rest}^{air}$ [\AA]} & \tablehead{1}{l}{b}{Branch} & \tablehead{1}{r}{b}{J} & \tablehead{1}{r}{b}{$\omega$} & \tablehead{1}{l}{b}{magn.~sen.} & \tablehead{1}{l}{b}{$\lambda_{\rm rest}^{air}$ [\AA]} & \tablehead{1}{l}{b}{Branch} & \tablehead{1}{r}{b}{J} & \tablehead{1}{r}{b}{$\omega$} & \tablehead{1}{l}{b}{magn.~sen.} \\ 
\hline
9950.34 & R & 10.5 & 1.5 & no & 9971.73 & R & 4.5 & 0.5 & no \\ 
9951.27 & P & 16.5 & 2.5-3.5 & yes & 9975.48 & Q & 2.5 & 2.5 & yes \\ 
9953.99 & R & 22.5 & 3.5 & yes & 9976.4 & R & 2.5 & 0.5 & yes \\ 
9954.59 & R & 12.5 & 1.5 & no & 9978.72 & R & 6.5 & 0.5 & no \\ 
9971.07 & R & 12.5 & 1.5 & no & 9979.87 & Q & 3.5 & 2.5 & yes \\ 
\hline
\end{tabular}
\caption{Several quantities of the investigated FeH lines
 \cite{2006ApJ...644..497R}. $J$ is the angular momentum and $\Omega$ the
 ro-vibrational constant. Magn.~sen. stands for magnetic sensitivity. }
\label{tab2}
\end{table}

To quantify the results visualized in Fig.~\ref{fig1}, we measured the
equivalent width, the gaussian FWHM, and the line depth of the ten
investigated FeH-lines (see Tab.~\ref{tab2}). A run of these quantities is
plotted in Fig.~2.\\
\textbf{Equivalent width (Fig.~2 top)}:
The equivalent width shows little difference between the 3D and the
<3D> lines. This difference stays almost constant in both cases (different
$\teff$ and $\log{g}$). This indicates that the deviations between the FeH
lines are due to velocity broadening, which should only mildly affect the equivalent
width. The run of the equivalent width in the $\log{g}$-series seems to stay
almost constant and is only weakly dependent on surface gravity. With
different effective temperatures the number of FeH molecules varies strongly
and hence the equivalent width. It decrease with increasing $\teff$ due to the
dissociation of the FeH-molecules. 

\textbf{Gaussian FWHM (Fig.~2 middle)}:
 The dependence of the line width (FWHM) on the surface gravity is very
 different for 3D- and <3D>-models. In the 3D case the FWHM decreases but in
 the <3D> case the FWHM increases with increasing surface gravity. The
 difference between 3D and <3D> lines at $\log{g}=3.0$ is around $1~{\rm
 km/s}$. We point out that in the <3D>-model we did not add any micro-
 or macro-turbulence. The increasing line width in the 3D case is a
 consequence of the hydrodynamic velocity fields which increase strongly with
 decreasing $\log{g}$. 
 In the $\teff$
 series, the effect of velocity related broadening is not as strong as in the
 $\log{g}$ case, but we believe that the difference between 3D- and
 <3D>-models stems from the appearing velocity fields, too.

\textbf{Line depth (Fig.~2 bottom)}:
The difference in line depths between 3D- and <3D>-models is consistent with the
velocity fields present in the atmospheres of the models. In the $\log{g}$
series the run of the line depth is very similar to the equivalent width and
reflects the almost constant number of FeH molecules. In the $\teff$ series, the line
depth shows a strong dependence on effective temperature and decreases with
increasing $\teff$. 

\begin{figure}[!h]
\resizebox{0.9\textwidth}{!}{
     \includegraphics[bb=60 230 550 630,clip]{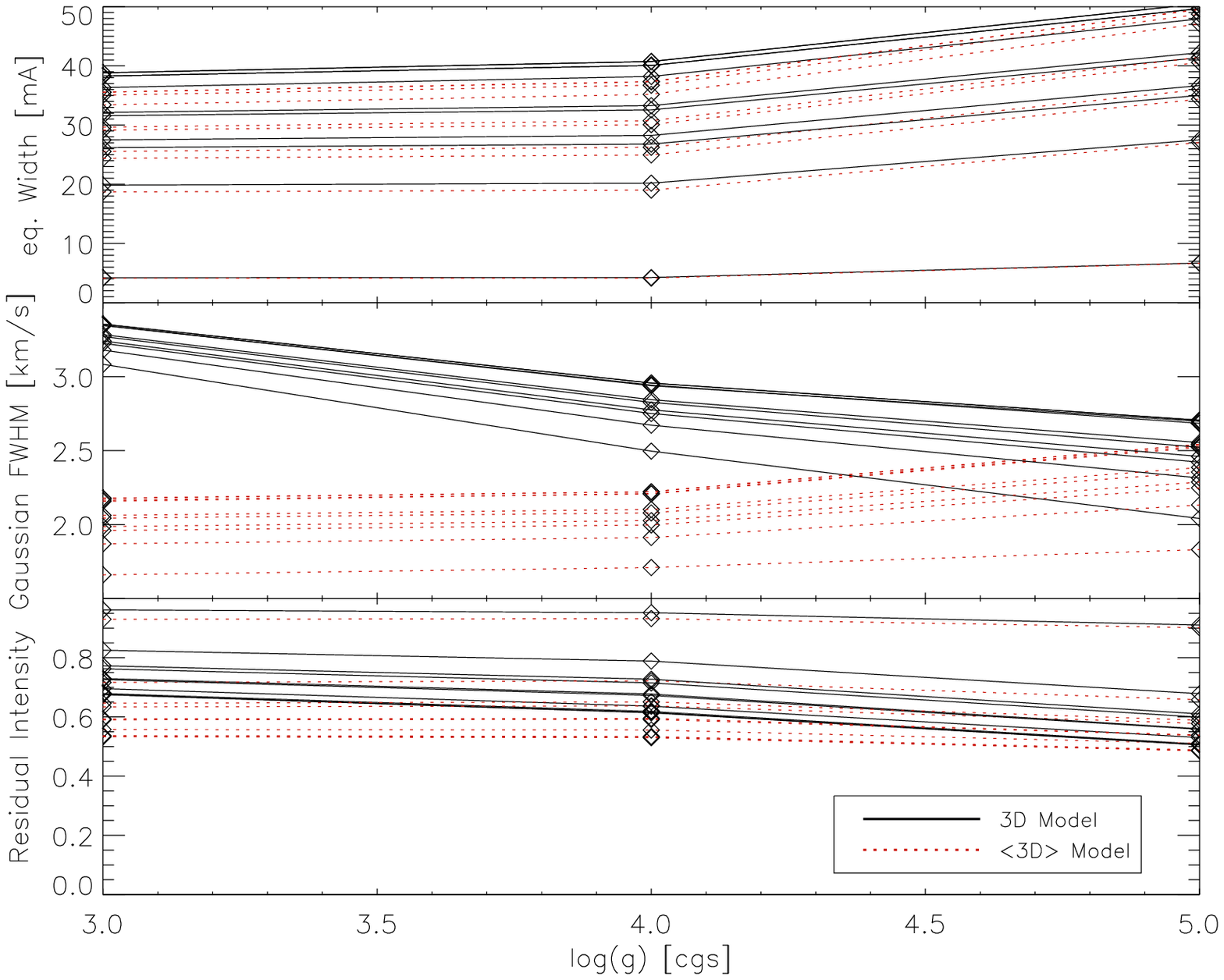}
     \includegraphics[bb=60 230 550 630,clip]{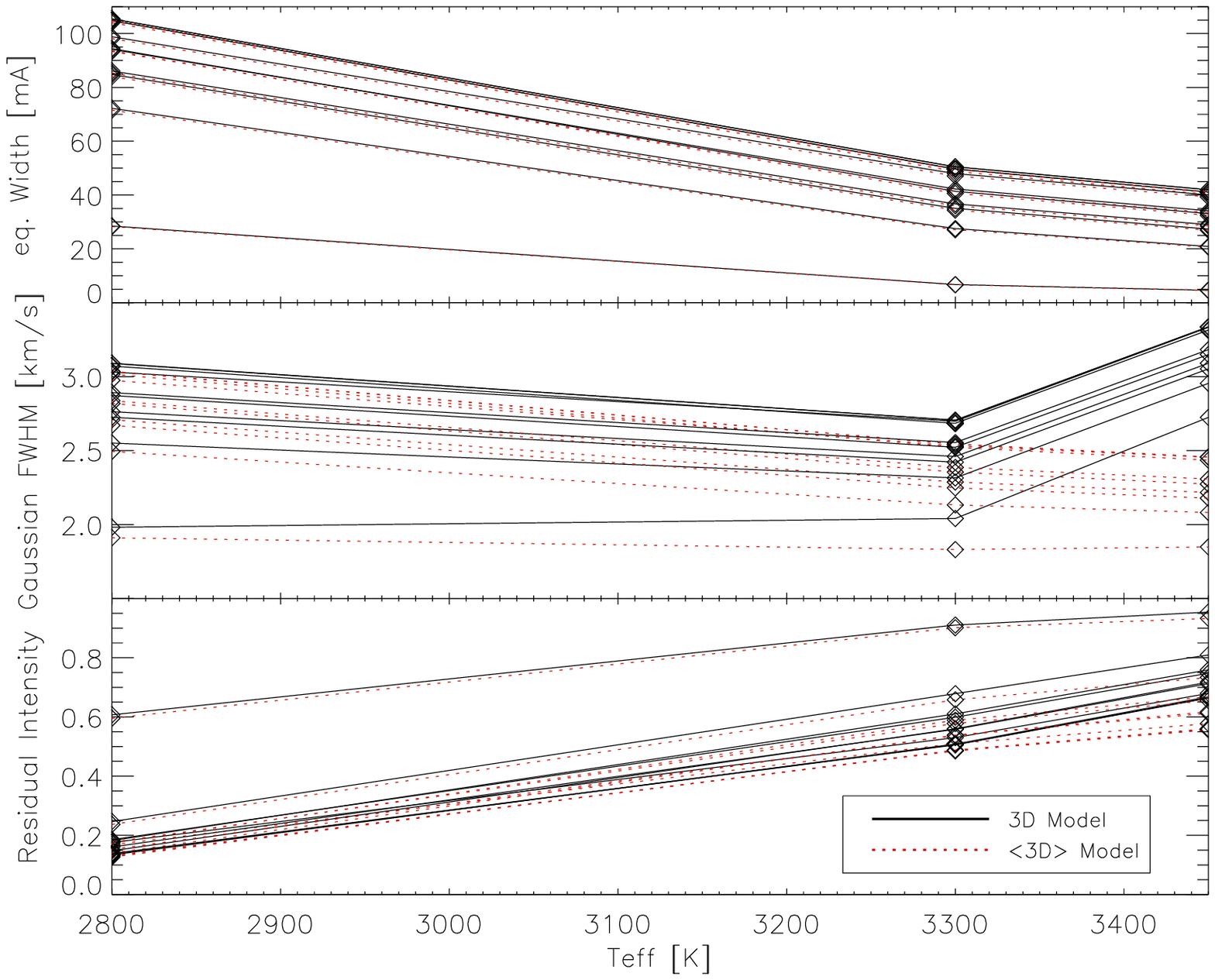}}
  \caption{\emph{Left:} Run of the equivalent width (top), the gaussian width
      (middle) and the line depth (bottom) for a model with $\teff=3300K$ and
      $\log{g}$ values of 3.0, 4.0 and 5.0 [cgs]. Plotted are the
      aforementioned quantities of the ten FeH-lines explained in
      Tab.\ref{tab2}. For more information see text.\hspace{6cm} \emph{Right:}
      Same quantities and lines for $\teff$'s of $2800~{\rm K}$, $3300~{\rm
      K}$ and $3450~{\rm K}$ and a $\log{g}$ value of 5.0 [cgs].}
\label{fig2}
\end{figure}

\section{velocity fields}
The velocity broadening has a noteworthy influence especially at low $\log{g}$
values. To investigate the velocity fields in the models, we use the
\emph{curve of growth} method. This employs the artificial increase of the
line strenght of a Ca\,I-line.  Ca\,I-lines produced on <3D>-models with
different micro-turbulent velocities enable us to determine a micro-turbulent
velocity for the 3D-model. We use the Ca\,I micro-turbulent velocities to
convolve the <3D>-model lines with gaussian velocity profiles. A comparison
with  <3D> lines which include the afore determined micro-turbulent velocities
computed with \verb!LINFOR3D!, shows that the difference between both lines is
insignificant. We broaden these
lines by convolution with a Gaussian profile again until they match the
3D lines and obtain an effective macro-turbulent velocity this way. The dependence
of the micro- and macro-turbulent velocities on surface gravity and effective
temperature is plotted in Fig.~3. One can see that the velocity field
represents the behavior of the quantities investigated above and gives an
explanation for the differences between 3D- and <3D>-models.

\begin{figure}[!h]
\resizebox{0.85\textwidth}{!}{
     \includegraphics{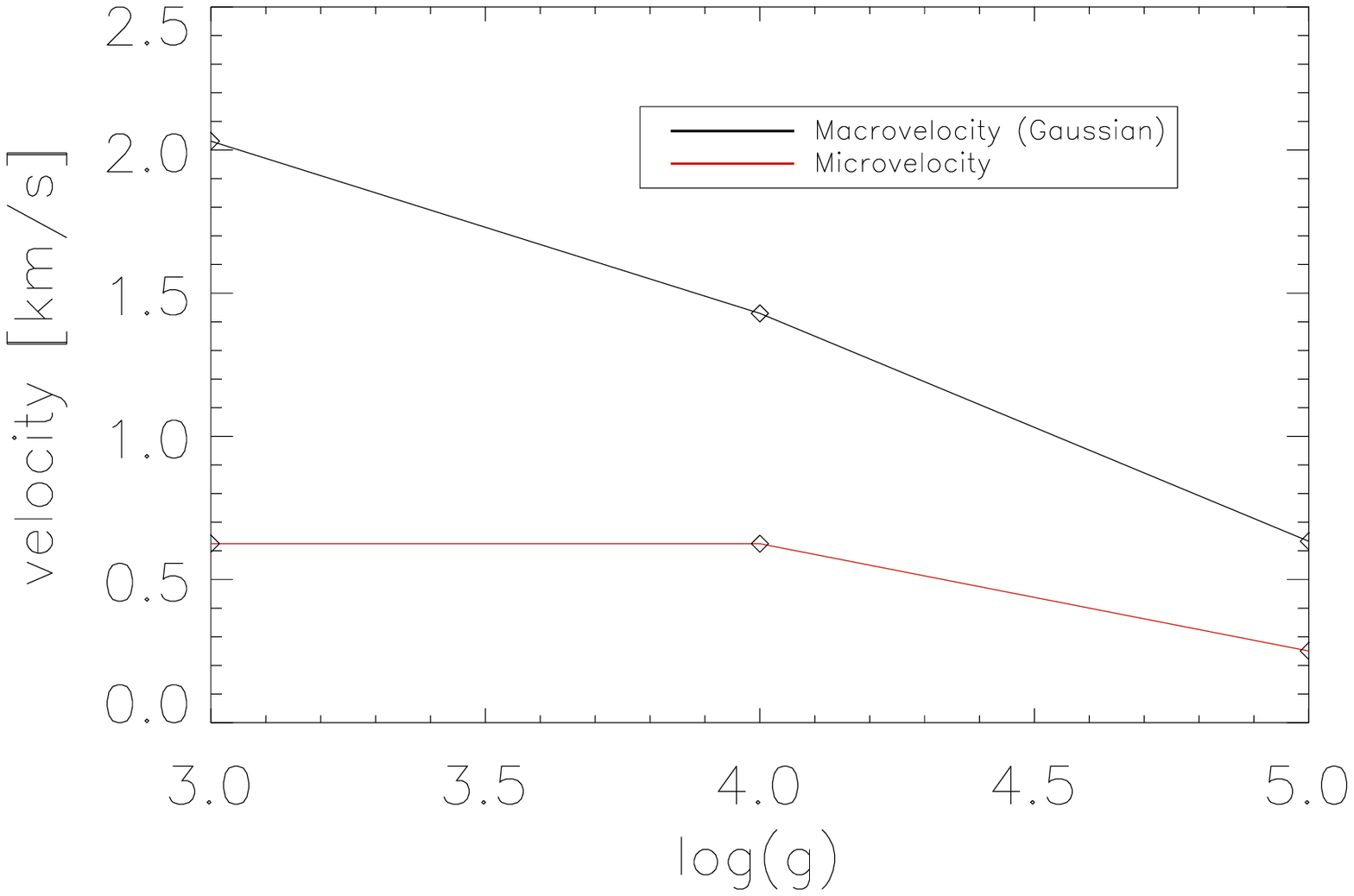}
     \includegraphics{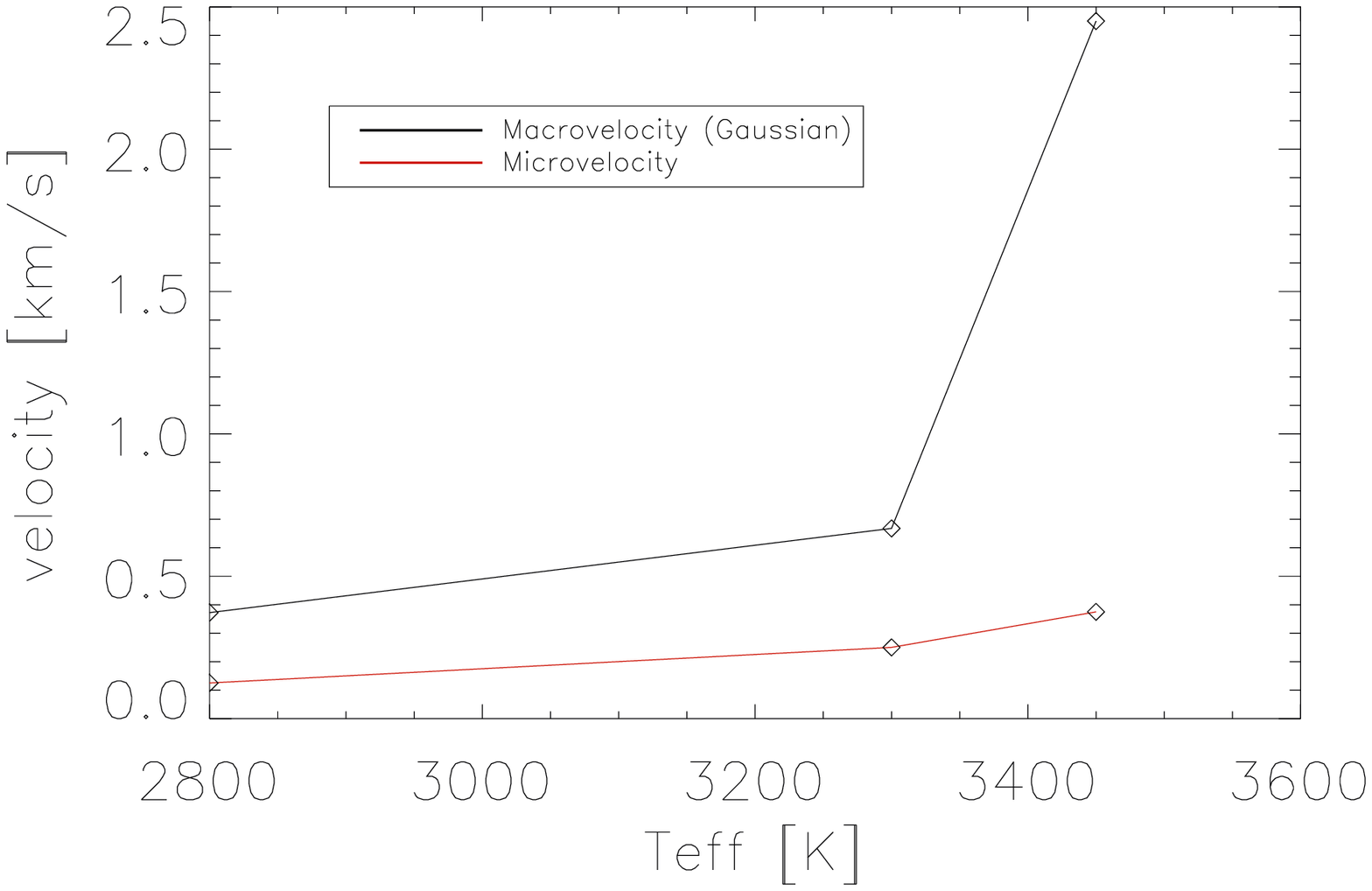}}
\caption{\emph{Left:} Micro- and macro-turbulent velocities as a function of
     $\log{g}$. \emph{Right:} Micro- and macro-turbulent velocities as a function of $\teff$.}
\label{fig3}
\end{figure}

\section{Discussion}
We have investigated ten single well isolated FeH lines between 9950~\AA\ and
9990~\AA\ on a set of 3D-\verb!CO5BOLD! models, with the spectral sythesis
code \verb!LINFOR3D!. We found that the FeH lines react on different effective
temperatures as one would expect due to the change in the FeH molecule
number. The lines also show a weak dependence on surface gravity but if one
includes the velocity-fields in the models, the $\log{g}$ dependence is strong
due to the strong velocity broadening. This means for the 1D spectral
synthesis that one has to include correct micro- and macro-turbulent
velocities for small surface gravities or the line width would be to small up
to $1~{\rm km/s}$. At high surface gravities, the FeH lines show a significant
difference between <3D> and 3D models only for higher effective
temperatures. This means that in the considered atmospheres thermal
Doppler broadening is the dominant velocity-related broadening mechanism.


\begin{theacknowledgments}
 SW would like to acknowledge the support from the DFG Research Training
Group GrK - 1351 ``Extrasolar Planets and their host stars''.
AR acknowledges research funding from the DFG under an Emmy Noether
Fellowship (RE 1664/4- 1).
HGL acknowledges financial support from EU contract MEXT-CT-2004-014265
(CIFIST).

\end{theacknowledgments}

\bibliographystyle{aipproc}  

\bibliography{ms}

\end{document}